\documentclass[orivec,envcountsame,runningheads]{llncs}
\usepackage{amsmath}
\usepackage{amssymb}
\newcommand{\IR}{\mathbb{R}}

\newcommand{\IZ}{\mathbb{Z}}
\newcommand{\IN}{\mathbb{N}}
\newcommand{\Ring}{\mathcal{R}}
\newcommand{\Field}{\mathcal{F}}
\newcommand{\calO}{\mathcal{O}}
\newcommand{\polylog}{\operatorname{polylog}}
\newcommand{\loglog}{\operatorname{loglog}}
\newcommand{\rev}{\operatorname{rev}}
\newcommand{\person}[1]{\textsc{#1}}
\newcommand{\aname}[1]{\textsf{#1}}
\newcommand{\mycite}[2]{{\rm\cite[\textsc{#1}]{#2}}}
\newcommand{\sgn}{\operatorname{sgn}}
\newcommand{\COMMENTED}[1]{}

\newtheorem{observation}[theorem]{Observation}
\newtheorem{fact}[theorem]{Fact}

\let\temp=\labelitemi
\let\labelitemi=\labelitemii
\let\labelitemii=\temp
\begin{document}
\title{Fast (Multi-)Evaluation of Linearly Recurrent Sequences:
Improvements and Applications}
\titlerunning{Fast (Multi-)Evaluation of Linearly Recurrent Sequences}
\author{Martin Ziegler\thanks{supported by DFG project \textsf{Zi1009/1-1}}}
\institute{PaSCo and HNI, University of Paderborn, 33095 GERMANY}
\maketitle
\begin{abstract}
For a linearly recurrent sequence
$\vec P_{n+1}=A(n)\cdot\vec P_n$,
consider the problem of calculating
either the $n$-th term $\vec P_n$
or $\ell\leq n$ arbitrary terms
$\vec P_{n_1},\ldots\vec P_{n_\ell}$,
both for the case of constant coefficients
$A(n)\equiv A$ and for a matrix $A(N)$
with entries polynomial in $N$.

We improve and extend known algorithms for
this problem and present new applications
for it. Specifically it turns out that for instance
\begin{itemize}
\item any family $(p_n)$ of classical orthogonal
  polynomials admits evaluation at given $x$
  within $\calO(\sqrt{n}\cdot\log n)$ operations
  \emph{independent} of the family $(p_n)$ under consideration.
\item For any $\ell$ indices $n_1,\ldots,n_\ell\leq n$,
  the values $p_{n_i}(x)$ can be calculated simultaneously
  using $\calO(\sqrt{n}\cdot\log n+\ell\cdot\log\tfrac{n}{\ell})$
  arithmetic operations; again this running time bound holds uniformly.
\item Every hypergeometric (or, more generally, holonomic)
  function admits approximate evaluation up to
  absolute error $\epsilon>0$ within
  $\calO(\sqrt{\log\tfrac{1}{\epsilon}}\cdot\loglog\tfrac{1}{\epsilon})$
  --- as opposed to $\calO(\log\tfrac{1}{\epsilon})$ ---
  arithmetic steps.
\item Given $m\in\IN$ and a polynomial $p$ of degree $d$
  over a field of characteristic zero,
  the coefficient of $p^m$ to term $X^n$
  can be computed within $\calO\big(d^2\cdot M(\sqrt{n})\big)$
  steps where $M(n)$ denotes the cost of multiplying
  two degree--$n$ polynomials.
\item The same time bound holds for the joint calculation
  of any $\ell\leq\sqrt{n}$ desired coefficients of $p^m$ 
  to terms $X^{n_i}$, $n_1,\ldots,n_\ell\leq n$.
\end{itemize}
\end{abstract}
\section{Introduction}
The naive way of calculating the $n$-th factorial
$P_{n+1}=(n+1)\cdot P_n$ uses
$\calO(n)$ arithmetic operations over $\IZ$.
During its course, all lower factorials 
$1,2,3!,\ldots,(n-1)!$ are generated
as well which might or might not be desirable.
In the latter case, most of the intermediate
factorials can in fact be bypassed and $n!$
itself be calculated using only
$\calO(\sqrt{n}\cdot\log n\cdot\loglog n)$ 
integer operations. 
This had been observed by
\person{Strassen} \mycite{Abschnitt~6}{Strassen}
and is based on
fast fourier transforms and polynomial multipoint
evaluation.
A generalization to the computation of the
$n$-th element $P_n$ of a recursively defined
sequence of vectors
\begin{equation} \label{e:VecDef}
{\vec P}_{n+1} \quad = \quad A(n)\;\cdot\;\vec P_n
\end{equation}
with a matrix $A$ of polynomials in $n$
has been suggested in \mycite{Section~6}{Chudnovsky},
further improved in \cite{Alin03} and
extended in \mycite{Theorem~5}{Alin04} to the joint computation 
of several (say, $\ell$, not necessarily consecutive) 
vectors $\vec P_{n_i}$.
This result has yielded better upper complexity bounds for
deterministic integer factorization and for computation
with hyperelliptic curves \cite{Alin03,Alin04}.

The present work reveals the \textsf{fast multi-evaluation
of linearly recurrent sequences} 
to be in fact fundamental for several other problems as well;
specifically to the evaluation of orthogonal polynomials and
to the computation of specific coefficients of very high
degree polynomials. Efficient handling of polynomials is
itself a basic ingredient to many fast algorithms with
a vast range of applications and, as a matter of fact,
plays in turn a major role in the fast evaluation of recurrent
sequences. 

We first review and extend the previously known algorithms
for linearly recurrent sequences with both constant and
with polynomial coefficients (Section~\ref{s:Recurrences}).
These are then applied to other problems as follows:
Section~\ref{s:Orthogonal} deduces
a roughly radial upper algebraic complexity bound 
\cite{ACT} uniformly on all orthogonal polynomials;
and Section~\ref{s:Partial} presents computer algebra
algorithms \cite{MCA2} for determining specific
coefficients of polynomials.
This is of particular benefit in cases where the result
has high degree $n$ but only few (say, $\ell\ll n$) 
terms are desired at output-sensitive cost.

In fact, all obtained running times are optimal with 
respect to $\ell$ in the following sense: As $\ell\to n$
(that is towards the classical case of computing all
entries of the sequence or coefficients of the polynomial)
and with other parameters fixed,
it converges to the asymptotic running time of the respective 
best known classical algorithm \emph{including}
logarithmic factors. Our new algorithms are thus true generalizations
of the latter at cost increased by at most constant factors.
\section{Fast Evaluation of Linear Recurrences} \label{s:Recurrences}
Recurrence equations like (\ref{e:VecDef}) are ubiquitous in
mathematics as well as computer science. Many (if not most) have 
no closed-form solution; and even if one does, it might
not induce an efficient algorithm --- compare $n!$ above.

In order to explicitly calculate the
$n$-th term $\vec P_n$, the naive approach suggested by Equation~(\ref{e:VecDef})
iteratively proceeds from $\vec P_0$ to $\vec P_1$, $\vec P_2$, \ldots, 
$\vec P_{n-1}, \vec P_n$
and thus has running time proportional to $n$.
However, being interested in $P_n$ only, this might be out-performed by 
other methods which avoid computing all intermediate terms.
For instance if $k\times k$--matrix $A$ does not depend on $N$, 
then repeated squaring yields $A^n$ within 
$\calO(k^3\cdot\log n)$ steps.
This is already optimal with respect to $n$ \mycite{Theorem~13.14}{ACT};
whereas in terms of $k$, the running time has been further improvemed
in \cite{Fiduccia} 
to time $\calO(k\cdot\polylog k\cdot\log n)$
for computing the solution to
\begin{equation} \label{e:ScalDef}
P_n \;\;=\;\; a_1\cdot P_{n-1} \;+\; a_2\cdot P_{n-2}
\;+\; \ldots \;+\; a_k\cdot P_{n-k} \enspace .
\end{equation}
Notice that (\ref{e:ScalDef}) is indeed a special case of
(\ref{e:VecDef}) with constant matrix $A$ of companion form
(\ref{e:Companion})
and $\vec P_n:=(P_n,P_{n-1},\ldots,P_{n-k+1})^\dagger$.
\begin{theorem} \label{t:Fiduccia}
Let $\Ring$ denote a commutative ring with 1
supporting multiplication of two polynomials
of degree $<n$ at cost at most $M(n)\geq n$.
\begin{enumerate}
\item[a)]
Given $a_1,\ldots,a_k\in\Ring$, $P_0,\ldots,P_{k-1}\in\Ring$, 
and $n\in\IN$, 
$\ell$ consecutive elements $P_{n},\ldots,P_{n+\ell-1}$ 
defined by (\ref{e:ScalDef})
can be computed using $\calO\big(M(k)\cdot\log n+\ell\cdot k\big)$
arithmetic operations in $\Ring$.
\item[b)]
Given a constant companion matrix $A\in\Ring^{k\times k}$ 
and $\vec P_0\in\Ring^k$, 
vectors $\vec P_{n_1},\vec P_{n_2}$, $\ldots$, $\vec P_{n_\ell}$ defined
by (\ref{e:VecDef}) can be computed simultaneously
using $\calO\big(\ell\cdot k^{\omega-1}\cdot\log\tfrac{nk}{\ell}\big)$
arithmetic operations in $\Ring$ where
$k\leq\ell$ and $n_1<n_2<\ldots<n_\ell=:n$.
\end{enumerate}
Here, $\omega\geq2$ denotes any feasible exponent for
matrix multiplication {\rm\cite{MCA2,ACT}}; e.g., $\omega=2.38$.
For $\ell\geq k^2$, one may even choose $\tilde\omega=2.34$.
\end{theorem}
\begin{proof}
Claim~a) for $\ell=1$ is \mycite{Proposition~3.2}{Fiduccia}.
Specifically, 
$P_n$ is obtained as the calar product of $(P_0,\ldots,P_{k-1})$
with the coefficients of the polynomial $X^n\bmod f$
where $f=X^k-(a_1+a_2X+\cdots+a_kX^{k-1})$
\mycite{Theorem~3.1}{Fiduccia}.
Therefore, once $X^n\bmod f$ is known, 
we can calculate $X^{n+1}\bmod f=\big(X^n\bmod f)\bmod f$
and $P_{n+1}$ using an additional number $\calO(k)$ of
operations. Iteration thus establishes the case $\ell>1$.

For Claim~b), use \mycite{Proposition~2.4}{Fiduccia} to compute
all binary powers of $A$ up to $n$ within $\calO\big(k^2\cdot\log n\big)$.
Therefore, each $\vec P_{n_i}$ is the product of $\vec P_0$ with
$J:=\calO(\log n)$ of these pre-calculated matrices $A^{2^j}$,
$j=1,\ldots,J$.
In order to improve the induced naive running time of 
$\calO(\ell\cdot k^2\cdot\log n)$
for the joint computation of $\vec P_{n_1},\ldots,\vec P_{n_\ell}$,
batch the matrix-vector products into matrix-matrix products as follows:
For each $j=0,\ldots,J$,
collect all $i=1,\ldots,\ell$ for which $\vec P_{n_i}$ involves $A^{2^j}$
in the above mentioned product; put the corresponding vectors 
to be multiplied to $A^{2^j}$ as columns into 
a $k\times\ell$ matrix and multiply that to $A^{2^j}$
using $\calO\big(k^\omega\cdot\big\lceil\tfrac{\ell}{k}\big\rceil\big)$ operations.
Here, $\omega=2.38$ is feasible due to \cite{Omega};
alternatively, one can use 
$\calO\big(k^{\hat\omega}\cdot\big\lceil\tfrac{\ell}{k^2}\big\rceil\big)$
operations with $\hat\omega=3.34$  \cite{Huang}.
Since $\ell\geq k$ (or $\ell\geq k^2$), this yields running time 
$\calO\big(\ell\cdot k^{\omega-1}\cdot\log n\big)$.

More careful analysis reveals that it suffices to multiply
only one vector (namely $\vec P_0$) to $A^{2^0}$ in time
$\calO(k^2)$; and two vectors (namely $\vec P_0$ and $A^{2^0}\cdot\vec P_0$)
to $A^{2^1}$ in twice the time; and, similarly on,
to multiply in phase no.$j$
only $2^j$ vectors to $A^{2^j}$ as long as
$j\leq\log(k)/\log(\omega-2)$ with $\calO(k^{\omega-2})$
vectors multiplied using a total of $\calO(k^{\omega})$ 
operations dominated by the last phase. 
From $j\geq\log(k)/\log(\omega-2)$ on, switch
to fast matrix multiplication. For $j\leq\log k$, this
involves only one bach and
will thus take time $\calO(k^{\omega})$ per phase, that is,
a total of $\calO\big(k^{\omega}\cdot\log k\big)$.
Each phase no.$j$ with $\log k\leq j\leq\log\ell$
gives rise to $2^j$ vectors multiplied to $A^{2^j}$,
grouped to $2^j/k$ batches and therefore 
taking $\calO\big(2^j\cdot k^{\omega-1}\big)$ operations,
again dominated by the last one with duration
$\calO\big(\ell\cdot k^{\omega-1}\big)$.
The final phases $j=\log\ell\ldots\log n$ do not further
increase the number of vectors multiplied to $A^{2^j}$
because we are looking for only $\ell$ different
results $\vec P_{n_1},\ldots,\vec P_{n_\ell}$.
They thus induce total cost 
$\calO\big(\ell\cdot k^{\omega-1}\big)$ each;
times the number $\log\tfrac{n}{\ell}$ of final phases
and added to the aforementioned
$\calO\big(k^{\omega}\cdot\log k\big)$
yields the claim.
\qed
\end{proof}
For further improvement and regarding the very last
paragraph of \cite{Fiduccia}, it seems worth while
to attack the following
\begin{problem} \label{p:Fiduccia}
Given $a_1,\ldots,a_k$ and $P_0,\ldots,P_{k-1}$,
~compute $P_{k},\ldots,P_{2k-1}$ according
to (\ref{e:ScalDef}) in time $o(k^2)$.
\end{problem}
Let us now relax the condition on $A$ to be constant
and consider matrices\ldots
\subsection{\ldots with Polynomial Coefficients} \label{s:Nonconst}
This case involves not matrix powers but
matrix factorials like $A(n)\cdot A(n-1)\cdots A(2)\cdot A(1)=:\prod_{j=1}^n A(j)$.
While the naive iterative approach leads to running time
proportional to $n$, \person{Chudnovsky\&Chudnovsky}
have improved that to cost roughly radical in $n$
\mycite{Section~6}{Chudnovsky}:
\begin{fact} \label{f:Chudnovsky}
Let $\Ring$ denote a commutative ring
permitting multiplication of two polynomials
of degree $<n$ at cost at most $M(n)\geq n$
where $M$ satisfies some standard
regularity conditions {\rm\cite[bottom of p.582]{Brent}}.
Consider a $k\times k$ matrix $A(N)$
with polynomial entries
$a_{ij}(N)\in\Ring[N]$ in $N$ of degree $<d$.
Given (the coefficients of) $A$ and $n\in\IN$,
one can calculate the matrix $\prod_{j=1}^n A(j)$ using
~$\displaystyle\calO\big(k^\omega\cdot M(\sqrt{n}d)\cdot\log n\big)$
operations in $\Ring$.
\end{fact}
\begin{proof} 
Let $\nu:=\lceil\sqrt{n}\rceil$ and consider the
Baby-Step/Giant-Step approach of
\begin{enumerate}
\item[i)] determining the (coefficients of the) polynomial matrix
 $C(N):=A(N+\nu)\cdot A(N+\nu-1)\cdots A(N+2)\cdot A(N+1)\in\Ring[N]^{k\times k}$;
\item[ii)] multi-evaluating $C$ at
  $0,\nu,2\nu,\ldots,\lfloor n/\nu\rfloor\cdot\nu=:\tilde n$;
\item[iii)] calculating
  $\prod_{j=1}^{\tilde n} A(j)$
  by iterative multiplication of the matrices
  ~$C(0), C(\nu)$, $\ldots, C(\tilde n-\nu)$~
  obtained in ii);
\item[iv)] finally computing $\prod_{j=1}^n A(j)$
  by iterative multiplication of the result from iii)
  with the matrices
  $A(\tilde n), A(\tilde n+1), \ldots, A(n-1)$.
\qed\end{enumerate}
\end{proof}
The possibility for further improvement of
Fact~\ref{f:Chudnovsky} to, say, $\calO(\polylog n)$ for fixed $(k,d)$
is unknown already in the case of the scalar factorial $n!$
and related to deep open class separation problems 
in complexity theory \cite{Cheng,Koiran}.
For rings of characteristic 0 and fixed $d$ however, 
improvements in particular in terms of the size $k$ of the matrix $A$
have been obtained by \person{Bostan\&Gaudry\&Schost}
\mycite{Theorem~5}{Alin04} as well as a generalization
to the simultaneous computation of 
$\ell\leq\calO\big(n^{{1}/{2}-\epsilon}\big)$
matrix factorials $\prod_{j=1}^{n_i} A(j)$, $i=1,\ldots,\ell$.

The present section reviews this result,
presented with a new proof 
and including in its analysis
the running time's dependence on the degree $d$ of the
polynomials in $A$ as well as on the number $\ell$
of elements of the sequence to be computed 
non-trivially extended beyond $\sqrt{n}$ 
(Theorem~\ref{t:Alin}b).
Further claims deal with a generalization 
(Theorem~\ref{t:Alin}a) and improvements
for the frequent case that
$A$ has companion form (Theorem~\ref{t:Alin}c+d).
In the sequel, capital letters $X$ and $N$
denote formal indeterminates of polynomials
whereas lower case $x$ and $n$ refer to
variables with values.
\begin{theorem} \label{t:Alin}
Consider a $k\times k$ matrix $A(N)$
with polynomial entries
$a_{ij}(N)\in\Ring[N]$ of degree $<d$.
\begin{enumerate}
\itemsep3pt
\item[a)]
  Given (the coefficients of) $A$ as well as
  $\ell$ pairs of integers
  $(m_i,n_i)$ with $0\leq m_i\leq n_i$,
  one can simultaneously calculate
  the $\ell$ matrix products
  $B_i:=\prod_{j=m_i}^{n_i} A(j)$, $i=1,\ldots\ell$, using 
  $$ \calO\Big(
     k^\omega\cdot\big(\sqrt{nd}+\ell\log\ell\big) \;+\; 
     k^2\cdot M\big(\sqrt{nd}\big) \;+\;
     k^2\cdot\ell\cdot\tfrac{M({nd}/{\ell})}{nd/\ell}\Big) $$
  operations in $\Ring$ where $n:=\max_i n_i\geq d\cdot\log^2 d$.
\item[b)]
  If $m_i\equiv1$ and, instead of the matrices $B_i$ themselves,
  the $\ell$ matrix--vector products $\vec P_i=B_i\cdot\vec P_0$
  for a given $\vec P_0\in\Ring^k$ are desired,
  this can be accomplished using
  $$  \calO\Big(k^\omega\cdot\min\big\{\sqrt{nd},{nd}/{\ell}\big\}\;+\;
                k^2\cdot M\big(\sqrt{nd}\big) \;+\;
                k^2\cdot\ell\cdot\tfrac{M({nd}/{\ell})}{nd/\ell}\Big) $$
  operations in $\Ring$.
\item[c)]
  In case that the matrix $A(n)$ is of companion form 
  and invertible in $\Ring^{k\times k}$ for
  all integers $n$ exceeding a given $m\in\IN$, 
  then the $\ell$ vectors $B_i\cdot\vec P_0$,
  $i=1,\ldots,\ell$, can be computed using
  $$ \calO\Big(k^2\cdot M\big(\sqrt{nd}\big) \;+\;
     k^2\cdot\ell\cdot\tfrac{M({nd}/{\ell})}{nd/\ell}\Big) $$
  operations in $\Ring$.
\item[d)]
  If additionally $n\geq k^2$ and the polynomials constituting $A(N)$
  obey the restricted degree condition
  $\deg(a_{1j})\leq j$, the running time further reduces to
  $$\calO\Big(k^2\cdot\big(\tfrac{\ell}{\sqrt{n}}+1\big)\cdot M(\sqrt{n})\Big)$$
\end{enumerate}
The algorithms
are uniform and --- except
for the roots of unity $\exp(2\pi i/n)$ employed
in the FFT when $M(n)=\calO(n\log n)$ --- free of constants.
\end{theorem}
\subsection{Proof of Theorem~\ref{t:Alin}}
Reconsider the proof of Fact~\ref{f:Chudnovsky}
with its four steps, but leave the value of the
trade-off parameter $\nu$ open for the moment
to be chosen later as an integral power of 2.
We also remark that the coefficients of the
polynomials arising in Steps~i) and ii)
may be taken with respect to any common
(rather than the standard monomial) basis.
As a matter of fact, regarding the hypothesis that $n\geq d\log^2d$, 
it pays off to first spend
$\calO\big(k^2\cdot M(d)\cdot\log d\big)$ operations for converting
$A(N)$ to the \aname{falling factorial}
(also called \aname{Newton}) basis
\mycite{Section~4.2}{Gerhard} because
that will accelerate evaluation and interpolation
on arithmetic progressions by a logarithmic factor
\mycite{Section~4.3}{Gerhard}. 
Specifically exploit that
evaluating a degree-$D$ polynomial $p$ simultaneously
at $K$ points of an arithmetic progression
takes, by simulating $\lceil K/D\rceil$ multipoint
evaluations of $p$ at $D=\deg(p)$ points each,
$\calO\big((\tfrac{K}{D}+1)\cdot M(D)\big)$
operations \mycite{Theorem~4.24}{Gerhard}.
Step~ii) thus succeeds within a total of
$\calO\big(k^2\cdot(\tfrac{n/\nu}{\nu d}+1)\cdot M(\nu d)\big)$
operations.

Concerning Step~i), \mycite{Section~6}{Chudnovsky} combines
fast matrix multiplication with fast polynomial arithmetic
and achieves running time 
$\calO\big(k^\omega\cdot M(\nu d)\log\nu\big)$.
\cite{Alin03} has observed that this allows for
improvement, provided the
characteristic of $\Ring$ is zero (or larger than $m+\nu d$).
Their proof is a recursive descend on $n$ being
an integral power of $4$ with a complicated
consideration for the general case.
We obtain a considerable simplification
in particular in Sub-Steps~$\alpha$) 
and $\gamma$) below
by working in the Newton rather than
monomial basis:
\begin{enumerate}
\item[$\alpha$)] Perform $k^2$ separate 
multipoint evaluations to obtain the matrix \emph{values}
$A(m+1),A(m+2),\ldots,A(m+2\nu d)\in\Ring^{k\times k}$
for arbitrary $m\in\IN$. Since the evaluation points
form an arithmetic progression this takes,
similarly to Step~ii), 
a total of $\calO\big(k^2\nu\cdot M(d)\big)$
operations. 
\item[$\beta$)] Determine the matrices
$C(m), C(m+1), \ldots, C(m+\nu d-1)\in\Ring^{k\times k}$
using $\calO(k^\omega\nu d)$ analogously to
\mycite{Proposition~2}{Alin03}.
Specifically, compute the $2\nu$ products 
$A(m+\nu)\cdot A(m+\nu-1)\cdots A(m+\nu-j+1)$
and $A(m+\nu+j)\cdot A(m+\nu+j-1)\cdots A(m+\nu+1)$
for $i=0,1,\ldots,\nu$ within
$\calO(k^\omega\nu)$
and observe that each $C(m),\ldots,C(m+\nu-1)$
is composed of two such product ranges.
$\big[C\big(m+\nu\big),\ldots,C\big(m+2\nu-1\big)\big],\ldots,
\big[C\big(m+(d-1)\nu\big),\ldots,C\big(m+d\nu-1\big)\big]$ 
are obtained similarly.
\item[$\gamma$)] Interpolate the $\nu d$ matrix values 
from Sub-Step~$\beta$) to determine
the (coefficients in the factorial basis of the)
matrix polynomial $C(N)$ of degree $<\nu d$
at the expense of $\calO\big(k^2\cdot M(\nu d)\big)$
operations \mycite{Theorem~4.26}{Gerhard}.
\end{enumerate}
Since $M(\nu d)\leq\nu M(d)$, 
the asymptotic cost of Step~ii) above exceeds that of
Sub-Step~$\alpha$), Step~i) gives rise to an additional
running time of $\calO\big(k^\omega\nu d+k^2 M(\nu d)\big)$.
Step~iii) uses $\calO\big((1+\tfrac{n}{\nu})\cdot k^\omega\big)$
operations 
and Step~iv) another $\calO(\nu\cdot k^\omega)$.

If, instead of the matrix $\prod_{j=1}^n A(j)$ itself,
only the vector $\vec P_n=\prod_{j=1}^n A(j)\cdot\vec P_0$
is to be calculated, we can replace the $\calO(k^\omega)$-time
matrix-matrix products in Steps~iii) and iv) with 
$\calO(k^2)$-time matrix-vector products.
If furthermore $A(n)$ is in companion form and
invertible for all integers $n\geq m$,
also Step~i$\beta$) accelerates to
$\calO(k^2\nu d)$ by Lemma~\ref{l:Companion}c) below.

\smallskip
Towards the multi-evaluation case $\ell>1$, 
suppose for a start that all $n_i$ and $m_i$ 
are multiples of $\nu$. We thus seek an algorithm 
for the following step:%
\vspace*{-1ex}\begin{enumerate}
\item[v)] Simultaneously calculate the $\ell$ matrices
~$\prod_{j=\tilde m_i+1}^{\tilde n_i} A(j)$~
(or their respective product with $\vec P_0$) where
$\tilde n_i:=\lfloor n_i/\nu\rfloor\cdot\nu$ and
$\tilde m_i:=\lceil m_i/\nu\rceil\cdot\nu$,
$i=1,\ldots,\ell$.
\vspace*{-1ex}\end{enumerate}
For Claims~b+c) with $\tilde m_i\equiv0$, 
it suffices to iteratively multiply the 
matrices $C(0),C(\nu),C(2\nu),\ldots$ obtained in ii):
this yields all products $\prod_{j=1}^{s\nu} A(j)$,
$s=1,\ldots,\rfloor n/\nu\rfloor=:I$ and takes
$\calO(\tfrac{n}{\nu}k^2)$ steps.
For Claim~a) with general $\tilde m_i$, we 
have to calculate $\ell$ products 
$\prod_{j=r_i}^{s_i-1}\tilde C_j$
of matrices $\tilde C_j:=C(j\nu)$ 
where the ranges $[r_i,s_i)$, $i=1,\ldots,\ell$ 
may be arbitrary integer intervals contained
in $[0,I)$. To this end recall the Range Tree 
from Computational Geometry \mycite{Section~5.1}{deBerg}. 
Specifically, consider the set
$S:=\{0,r_i,s_i:i=1,\ldots,\ell\}$
ordered as $S=\{0=t_0<t_1<\ldots<t_{\tilde N-1}\}$
where $\tilde N\leq\min\{2\ell+1,I\}$. Now compute first 
the $\tilde N$ products $\prod_{t\in[t_j,t_{j+1})}\tilde C_t$,
$j=0,\ldots\tilde N-1$, invoking
$\calO(\sum_{j} |t_{j+1}-t_j-1|)=\calO(I)$ matrix multiplications;
then compose from these results the $\tilde N/2$ products
$\prod_{t\in[t_{2j},t_{2j+2})}\tilde C_t$, $j=0,\ldots,\tilde N/2-1$
using further $\tilde N/2\leq I/2$ matrix multiplications;
then the $\tilde N/4$ products $\prod_{t\in[t_{4j},t_{4j+4})}\tilde C_t$,
and so on.
So after a total of $\calO(k^\omega I)$ operations, all
products ranging over a binary interval are prepared
which concludes the initialization of the Range Tree.
Now for its application,
observe that each interval $[r_i,s_i)$, $i=1,\ldots,\ell$ 
is a disjoint
union of $\calO(\log\tilde N)\leq\calO(\log\ell)$ of these
binary intervals. 
This concludes the entire Step~v) within time
$\calO\big(k^\omega(\tfrac{n}{\nu}+\ell\log\ell)\big)$
in case of Claim~a) or
$\calO(k^2\tfrac{n}{\nu})$ for Claims~b+c).

\smallskip
For the final goal, that is to
\vspace*{-1ex}\begin{enumerate}
\item[vi)] simultaneously calculate the $\ell$ matrices
~$\prod_{j=m_i}^{n_i} A(j)$~
(or their respective product with $\vec P_0$),
$i=1,\ldots,\ell$,
\vspace*{-1ex}\end{enumerate}
invoke the Range-Tree idea once again. This time, 
the initialization phase consists in preparing the
(coefficients of the) $\nu/2$ polynomial matrices
$C_{\nu/2}(N):=A(N+\tfrac{\nu}{2})\cdot 
   A(N+\tfrac{\nu}{2}-1)\cdots A(N+2)\cdot A(N+1)\in\Ring[N]^{k\times k}$,
$C_{\nu/4}(N)$, $C_{\nu/8}(N)$, \ldots, $C_{2}(N)$, $C_{1}(N)$.
Due to the exponentially decreasing size $\nu$, this will 
together infer only the same cost as Step~i).

In the application phase, first multi-evaluate $C_{\nu/2}(N)$ 
at those $\tilde n_i$ whose difference to $n_i$ is 
at least $\nu/2$ --- 
$\calO\big(k^2\cdot(\tfrac{\ell}{\nu d}+1)\cdot M(\nu d)\big)$
operations as in Step~ii) --- and multiply them to the already
computed results from Step~v) at the expense of another
$\calO(k^\omega\cdot\ell)$ and $\calO(k^2\cdot\ell)$ 
for Claims~a) and b+c), respectively.
For Claim~a) do similarly for those $\tilde m_i$ 
differing from $m_i$ by at least $\nu/2$.
Now repeat with multi-evaluating $C_{\nu/4}(N)$,
then $C_{\nu/8}(N)$ and so on.
By the same argument as above, this will affect the
overall running time by at most a factor of 2 while
in the end yielding the desired resulting values.

\begin{figure}[htb]
\renewcommand{\arraystretch}{1.3}
\begin{tabular}{r||l|l|l|l}
  & \hfill Case~a) \hspace*{\fill} & 
\hfill Case~b) \hspace*{\fill} &
\hfill Case~c) \hspace*{\fill} &
\hfill Case~d) \hspace*{\fill} \\
\hline
i) & $k^\omega\nu d+ k^2 M(\nu d)$
   & $k^\omega\nu d+ k^2 M(\nu d)$
   & $k^2\nu d+ k^2 M(\nu d)$ 
   & $k^2M(\nu+k)$ \\
ii)& $k^2\big(\tfrac{n/\nu}{\nu d}+1\big)M(\nu d)$
   & $k^2\big(\tfrac{n/\nu}{\nu d}+1\big)M(\nu d)$
   & $k^2\big(\tfrac{n/\nu}{\nu d}+1\big)M(\nu d)$
   & $k^2\big(\tfrac{n/\nu}{\nu+k}+1\big)M(\nu+k)$\\[-0.5ex]
iii)\\[-2.8ex]\raisebox{-1ex}{\!\!\!\!\!+iv)}%
    & $k^\omega\big(1+\tfrac{n}{\nu}+\nu\big)$
    & $k^2\big(1+\tfrac{n}{\nu}+\nu\big)$
    & $k^2\big(1+\tfrac{n}{\nu}+\nu\big)$
    & $k^2\big(1+\tfrac{n}{\nu}+\nu\big)$\\
v)  & $k^\omega\big(\tfrac{n}{\nu}+\ell\log\ell\big)$
    & $k^2\tfrac{n}{\nu}$
    & $k^2\tfrac{n}{\nu}$
    & $k^2\tfrac{n}{\nu}$ \\
vi) & $k^2\big(\tfrac{\ell}{\nu d}+1\big)M(\nu d)$
    & $k^2\big(\tfrac{\ell}{\nu d}+1\big)M(\nu d)$
    & $k^2\big(\tfrac{\ell}{\nu d}+1\big)M(\nu d)$
    & $k^2\big(\tfrac{\ell}{\nu+k}+1\big)M(\nu+k)$
\end{tabular}
\caption{\label{f:Alin}Big-Oh running times 
of steps i) to vi) 
in cases a) to d)}
\end{figure}
It thus remains to confirm that the costs of the above steps
i) to vi)
are all covered by the running times claimed in a), b), 
and c). To this end, choose 
$\nu$ as (an integral power of 2 close to but not exceeding) 
$\sqrt{n/d}$ if $\ell\leq\sqrt{nd}$
and $\nu:=n/\ell$ otherwise.
We remark that $\nu\leq\sqrt{n/d}$
holds in both cases, so 
$$\tfrac{n}{\nu} \;+\; 
 \big(\tfrac{n}{\nu^2d}+\tfrac{\ell}{\nu d}+1\big)\cdot M(\nu d) 
 \quad\leq\quad 
 \calO\big((\tfrac{n}{\nu}+\ell)\cdot\tfrac{M(\nu d)}{\nu d}\big)$$
which amounts to $\calO\big(M(\sqrt{nd})\big)$ in case 
$\ell\leq\sqrt{nd}$ and to 
$\calO\big(\tfrac{\ell^2}{nd}\cdot M(\tfrac{nd}{\ell})\big)$
if $\ell\geq\sqrt{nd}$; Claims~a+c) are thus immediate.
For Claim~b) observe furthermore that 
$\nu d=\min\{\sqrt{nd},{nd}/{\ell}\}$.

Case~d) admits, in addition to Case~c), further improvement
based on the observation that the degree of
the matrix polynomial(s) $C(N)$ involved in Steps~i), ii), and vi) 
reduces from $\nu d$ to $\nu+k$ by virtue of Observation~\ref{o:DegPattern}b)
below. This yields the running times in the last column
of Figure~\ref{f:Alin}.
Then choose $\nu:=\sqrt{n}\geq k$.
\qed
\subsection{A First Application and Some Tools}
Theorem~\ref{t:Alin}b) includes
\mycite{Theorem~5}{Alin04} by restricting to
$\ell\leq\calO(\sqrt{n})$ and constant $d$.
Another consequence, we have the following 
non-trivial complexity interpolation between,
on the one end, \person{Strassen}'s aforementioned
algorithm \mycite{Abschnitt~6}{Strassen} computing one
single factorial $n!$ (that is, the case $\ell=1$)
and, on the other end, 
the obviously optimal naive iterative
$\calO(n)$ calculation of 1, 2!, 3!, \ldots, $n!$
(that is, the case $\ell=n$):
\begin{corollary} \label{c:Factorial}
Over
$\Ring=\IZ$ with $M(n)=\calO(n\log n\loglog n)$ \mycite{Theorem~8.23}{MCA2}
and $k=1=d$,
any $\ell$ desired factorials $n_1!<n_2!<\ldots<n_\ell!=n!$
can be computed simultaneously using
$\calO\big(\sqrt{n}\cdot\log n\cdot\loglog n
+\ell\cdot\log\tfrac{n}{\ell}\cdot\loglog\tfrac{n}{\ell}\big)$
arithmetic operations.
\end{corollary}
Further applications will be given in the sequel.
In many of them, the matrix $A$ according to
Equation~(\ref{e:VecDef}) is structured \cite{Pan}.
For example a companion matrix as well as its inverse 
\begin{equation} \label{e:Companion}
F\;=\;\left(\begin{array}{cccccc}
f_1 & f_2 & f_3 & \ldots & f_{k-1} & f_k \\
1 & 0 & 0 & \ldots & 0 & 0 \\
0 & 1 & 0 & \ldots & 0 & 0 \\
0 & 0 & 1 & \ldots & 0 & 0 \\
\vdots & \vdots & & \ddots & \vdots & \vdots \\
0 & 0 & 0 & \ldots & 1 & 0 \end{array}\right), \qquad
F^{-1}\;=\;\left(\begin{array}{ccccc}
0 & 1 & 0 & \ldots & 0 \\
0 & 0 & 1 & \ldots & 0 \\
\vdots & \vdots & & \ddots & \vdots  \\
0 & 0 & 0 & \ldots & 1 \\
\tfrac{1}{f_k} & \:\tfrac{-f_1}{f_k} & 
\:\tfrac{-f_2}{f_k} & \ldots & \:\tfrac{-f_{k-1}}{f_k} 
\end{array}\right)
\end{equation}
is described by the $k$ parameters
$(f_1,\ldots,f_k)$ as opposed to the $k^2$ independent
entries of a general matrix. 
Theorem~\ref{t:Fiduccia} relies on fast powering of
companion matrices, that is, on efficient calculation 
of iterated products of the same $F$.
The following tool, employed in the proof of
Theorem~\ref{t:Alin}c), considers products
of several companion matrices and might be of its own interest:
\begin{lemma} \label{l:Companion}
Given $m$ companion matrices $F_1,\ldots,F_m\in\Ring^{k\times k}$,
\begin{enumerate}
\item[a)]
their product $F_1\cdots F_m$ can be computed in
$\calO(m\cdot k^2)$ steps
\item[b)]
as well as in $\calO\big(k^{\omega}\cdot(1+\tfrac{m}{k})\big)$ steps.
\item[c)]
If all $F_1,\ldots,F_m$ are invertible,
then the $m-n+1$ products
$$ F_1\cdots F_n, \quad F_2\cdots F_{n+1}, \quad\ldots,\quad F_{m-n+1}\cdots F_m $$
can be computed simultaneously 
in $\calO(m\cdot k^2)$ 
steps.
\end{enumerate}
\end{lemma}
\begin{proof}
\begin{enumerate}
\item[a)]
The multiplication of a vector to a companion matrix,
from left $F\cdot\vec v$ as well as its transposed
from left $\vec v^\dagger\cdot F$,
both takes $\calO(k)$ operations. 
Therefore the multiplication $A\cdot F_m$ by an
arbitrary square matrix like $A=F_1\cdots F_{m-1}$ takes $\calO(k^2)$ steps.
Iterating establishes the sought $\calO(m\cdot k^2)$ algorithm. 
\item[c)]
Compute the first product $P_1:=F_1\cdots F_n$ as in a).
For the subsequent terms
$P_{j+1} := F_{j+1}\cdots F_{n+j} = F_j^{-1}\cdot P_j\cdot F_{n+j}$
exploit as in a) that multiplication by $F_{n+j}$
as well as by $F_j^{-1}$ takes only $\calO(k^2)$ steps.
\item[b)]
Recall \mycite{Lemma~3.1}{Key} the formula
$$ F_1\cdot F_2\cdots F_{k-1}\cdot F_k \quad=\quad (I-L)^{-1}\cdot R $$
where $R$ and $L$ denote (respectively lower and strictly upper triangular) 
matrices plainly consisting of the $k^2$ joint parameters of 
$F_1,\ldots,F_k$. Both multiplication with $R$ and the inverse
$(I-L)^{-1}$ are feasible within $\calO(k^\omega)$ 
\mycite{Proposition~16.6}{ACT}. This establishes the case $m=k$; 
the general case now follows
by partitioning $m$ into $\lceil m/k\rceil$ blocks
of length $k$ each according to the grouping
$(F_{1}\cdots F_{k})\cdot(F_{k+1}\cdots F_{2k})\cdots\cdots(F_{m-k+1}\cdots F_{m})$.
\qed\end{enumerate}\end{proof}
The following improvement to Lemma~\ref{l:Companion} seems
conceivable:
\begin{problem} \label{p:Companion}
Given $k$ companion matrices of size $k\times k$,
compute their product using $\calO(k^2\cdot\polylog k)$
operations.
\end{problem}
Another ingredient to the proof of Theorem~\ref{t:Alin}d) is the following
\begin{observation} \label{o:DegPattern}
\begin{enumerate}
\item[a)]
Let $A\in\Ring[X]^{k\times k}$ and $\vec b\in\Ring[X]^k$
denote a matrix and vector of polynomials
of $\deg(b_j)\leq m-j$ and $\deg(a_{ij})\leq1+j-i$.
with the convention $\deg(0)=-\infty$.
Then, $\vec c:=F\cdot\vec b$ has 
$\deg(c_j)\leq m+1-j$.
\item[b)]
Let $F_1,\ldots,F_m\in\Ring[X]^{k\times k}$ denote
polynomial companion matrices with 
$(f_{i1},f_{i2}$, $\ldots,f_{ik})$ the first row of $F_i$, respectively.
If $\deg(f_{ij})\leq j$, then $B:=\prod_{\ell=1}^m F_{\ell}$
has $\deg(a_{ij})\leq m+j-i$.
\end{enumerate}
\end{observation}
$$ 
\deg(A)\;=\left(\begin{array}{ccccc}
1 & 2 & \ldots & k-1 & k \\
0 & 1 & \ldots & k-2 & k-1 \\
\vdots &  & \ddots & \vdots & \vdots \\
0 & 0 & \ldots & 1 & 2 \\
0 & 0 & \ldots & 0 & 1 \end{array}\right)
, \qquad
\deg(\vec b) \;=\; \left(\begin{array}{c}
m-1 \\ m-2 \\ \vdots \\ m-k+1 \\ m-k\end{array}\right)
$$
\begin{proof}
\begin{enumerate}
\item[a)] is a straight-forward consequence from $\deg(p\cdot q)\leq\deg(p)+\deg(q)$.
\item[b)] follows by induction on $m$,
applying a) to each column $\vec b$ of $B$.
\qed\end{enumerate}\end{proof}
\section{Fast Evaluation of Orthogonal Polynomials} \label{s:Orthogonal}
This sections concludes from Theorem~\ref{t:Alin}c) that
any family of classical orthogonal polynomials has 
at most roughly radical complexity\footnote{%
This is not to be confused with the 
Paterson\&Stockmeyer Result \cite{Paterson}
that \emph{every}
polynomial has complexity $\calO(\sqrt{n})$ 
when \emph{neglecting} operations in the
coefficient field.%
} $\calO(\sqrt{n}\cdot\log n)$.

\aname{Horner's Method} evaluates a fixed degree-$n$ polynomial
$p\in\Ring[X]$ at given $x$ within $\calO(n)$ arithmetic steps.
While this is optimal in the `generic' case
\mycite{Corollary~5.11}{ACT},
many specific polynomials do admit faster evaluation;
monomials $X^n$ for instance in time $\calO(\log n)$
by means of repeated squaring. Also 
Chebyshev's Polynomials $T_n\in\IZ[X]$ have 
complexity logarithmic in their degree;
this can be seen either directly from the quadratic recurrence
$$  T_{2n}(X) \;=\; 2 T_n^2(X)\,-\,1, \quad
    T_{2n+1}(X) \;=\; 2 T_{n+1}(X)\cdot T_n(X) \,-\, X $$
or by applying Theorem~\ref{t:Fiduccia} to the linear vector recurrence
$$ \binom{T_{n+1}(X)}{T_n(X)}  \;=\; A\cdot
  \binom{T_n(X)}{T_{n-1}(X)}, \qquad A:=
\bigg(\begin{array}{c@{\;\;\;}c} 2x & -1 \\[0.3ex] 1 & 0 \end{array}\bigg)
$$
with matrix $A$ independent of $n$
\mycite{Section~4}{Fateman}, \cite{KoepfCheby}.
Recall that $(T_n)$ forms an orthogonal system
on $[-1,+1]$ with respect to the weight $\rho(x)=(1-x^2)^{-{1}/{2}}$.
Other weights lead to other families of orthogonal polynomials.
They are a important tool in Mathematical Physics due to their
approximation properties \cite{Nikiforov}. The
\aname{Legendre Polynomials} $P_n(X)$ for instance
are orthogonal on $[-1,+1]$ with respect
to $\rho(x)\equiv1$. 
\begin{theorem} \label{t:Orthogonal}
Every monic family $(P_n)\subseteq\IR[X]$ of classical
orthogonal polynomials has complexity
$\calO(\sqrt{n}\cdot\log n)$.
Any $\ell$ members $P_{n_1},\ldots,P_{n_\ell}$ of
such a family have joint complexity 
$\calO(\sqrt{n}\cdot\log n\,+\,\ell\cdot\log\tfrac{n}{\ell})$.
The constants in the big-Oh notation are
\emph{independent} of the family $(P_n)$.
\end{theorem}
Observe that, as $\ell\to n$ (that is concerning the
problem of evaluating \emph{all} polynomials
$P_1(x),\ldots,P_n(x)$), the running time converges
to $\calO(n)$ which is clearly optimal.
\begin{proof}
It is well-known that any family $(P_n)$
of classical orthogonal polynomials
satisfies a three-term recursion
\begin{equation} \label{e:Orthogonal}
P_{n+1}(X) \;=\; (A_n\cdot X+B_n)\cdot P_n(X) \,-\, C_n\cdot P_{n-1}(X)
\end{equation}
see, e.g., \mycite{Section~II.6.3}{Nikiforov}.
In fact for monic $(P_n)$,
$A_n,B_n,C_n$ have turned out as rational functions
of $n$ with respective numerator and denominator polynomials
$a(N),b(N),c(N),\alpha(N),\beta(N),\gamma(N)\in\IR[N]$
of degree at most 4 \mycite{Theorem~1}{KoepfRecurrence}. 
Rewriting Equation~(\ref{e:Orthogonal}), we obtain
$$ \alpha(n)\beta(n)\gamma(n)\cdot
  \binom{P_{n+1}(X)}{P_n(X)}  =
\Bigg(\begin{array}{c@{\;\;\;}c} 
\displaystyle a(n)\beta(n)\gamma(n)\cdot X &
\displaystyle -\alpha(n)\beta(n)c(n) \\[1ex]
\displaystyle \alpha(n)\beta(n)\gamma(n) & 
\displaystyle 0 \end{array}\Bigg)
\cdot\binom{P_n(X)}{P_{n-1}(X)} $$
a recursion with polynomial coefficients of size $k$ and degree $d$
independent of the family $(P_n)$ under consideration.
Now apply Lemma~\ref{l:Polynomial} below with
$M(n)=\calO(n\cdot\log n)$.
\qed\end{proof}

\begin{lemma} \label{l:Polynomial}
Let $\Field$ denote a field of characteristic $0$
permitting multiplication of two polynomials
of degree $<n$ at cost at most $M(n)$.
Let $(\vec P_n)\subseteq\Field[X]^k$ be a sequence of
polynomial vectors satisfying
\begin{equation} \label{e:Polynomial}
s(n+1,X)\cdot \vec P_{n+1}(X) \;=\; A(n,X) \cdot \vec P_n(X)
\end{equation}
with companion matrix polynomial $A\in\Field[N,X]^{k\times k}$ 
and $s\in\Field[N,X]$ both of (total) degree $<d$.
Finally suppose that $s(n,x)\not=0$ for all $n\in\IN$ 
and all $x\in\tilde\Field$,
the latter denoting an arbitrary subset of $\Field$.
\\
Given $x\in\tilde\Field$, $\vec P_0(x)\in\Field^k$,
and (the order $k^2d^2$ coefficients of)
both $A$ and $s$, one can simultaneously evaluate
$\vec P_{n_1}(x),\ldots,\vec P_{n_\ell}(x)$ using
$$ \calO\Big(
    k^2\cdot M\big(\sqrt{nd}\big) \;+\;
    k^2\cdot\ell\cdot\tfrac{M({nd}/{\ell})}{nd/\ell}\Big) $$
arithmetic operations over $\Field$ where $\max\{d^3,n_i\}\leq n$.
\end{lemma}
The multi-evaluation expressed above
refers to the indices $n_1,\ldots,n_\ell$ of the sequence
and should not be confused with multipoint evaluation
of a polynomial at several point $x_1,\ldots,x_n$ as,
e.g., in \cite{Alpert}.
\begin{proof}[Lemma~\ref{l:Polynomial}]
Let $\sigma_n(X):=\prod_{i=1}^n s(n,X)$ and
consider the sequence $\vec Q_n:=\sigma_n\cdot\vec P_n$
obviously satisfying $\vec Q_{n+1}(X)=A(n,X)\cdot\vec Q_n(X)$.
After plugging in $x$ into $A$ using $\calO(k^2 d^2)$
arithmetic operations, one arrives thus in the situation
of Theorem~\ref{t:Alin}c). Indeed, if $a_{1,k}(N,x)\in\Field[N]$ 
is the zero polynomial, then we may truncate both the last
column and row of $A$ and reduce the dimension $k$ of the
recurrence by one; whereas if $a_{1,k}(N,x)$ is not identically
zero, it has only finitely many roots and $A(n,x)$ is invertible
for all $n\geq m$ with some appropriate $m$ which can easily
be found using standard bounds.
This yields the joint computation of
$\vec Q_{n_1}(x),\ldots,\vec Q_{n_\ell}(x)$
within the claimed time.
Now exploit $\sigma_{n+1}(X)=s(n+1,X)\cdot\sigma_n(X)$ to
similarly compute $\sigma_{n_1}(x),\ldots,\sigma_{n_\ell}(x)$.
Since these are units by assumption,
another $k\ell$ divisions yield the desired
values $\vec P_{n_i}(x)=\vec Q_{n_i}(x)/\sigma_{n_i}(x)$.
\qed\end{proof}
\section{Fast Partial Polynomial Arithmetic} \label{s:Partial}
The present section applies fast evaluation of linearly
recurrent sequences to the problem of
computing single or few specific coefficients of a polynomial
of large degree. 

Based on FFT-methods, many algorithms have been devised which yield
fast solutions to many problems in polynomial arithmetic 
\mycite{Part~II}{MCA2}.
These tend to be optimal in running time up to poly-logarithmic
factors, simply by comparison with the sizes of the input and output. 
However the operations of
\begin{description}
\item[composition:] given $p,q\in\Field[X]$, \quad determine $p\circ q$;
\item[powering:]  given $p\in\Field[X]$ and $n\in\IN$, \quad determine $p^n$;
\item[inversion:] given $p\in\Field[X]$ with $p(0)\not=0$ and $n\in\IN$, \\
  determine $q:=1/p\bmod X^n\in\Field[X]$.
\end{description}
generate results of degree significantly larger than the input:
quadratic in the first case, unbounded\footnote{We refer to
the \emph{algebraic} size of course; in terms of the \emph{bit} size
of $n$, the output is of exponential degree --- still too large.} 
in the second and third.
This leaves room for improved algorithms in cases where only 
one or few terms of the power or inverse are desired
--- preferably with output-sensitive running times 
proportional to the number of terms desired.
For instance, \mycite{Corollary~2.3}{ACT} accelerates polynomial
multiplication when some coefficients of the result are already
known. Our interest lies in situations where coefficients are not 
known nor of interest anyway, that is, in the \emph{partial}
calculation of polynomials. In this spirit, 
\cite{Brent} presents improved algorithms for computing
the lowest $\ell$ coefficients of the result where $\ell$ coincides with
the degree $d$ of the input \mycite{Corollary~2.33, Theorem~2.34}{ACT},
for \aname{composition} for instance in time
$\calO(d^{3/2}\cdot\polylog d)$.
Our result deals with determining either the $\ell$ 
most significant as well as arbitrary coefficients.
\begin{theorem} \label{t:Power}
Let $\Field$ denote a field 
permitting multiplication of two polynomials
of degree $<n$ at cost at most $M(n)$.
\begin{enumerate}
\item[a)]
  Given $p\in\Field[X]$ of degree $d$ and $n\in\IN$,
  the $\ell\geq d$ most significant coefficients of the power $p^n\in\Field[X]$
  can be computed in time $\calO\big(M(\ell)+\log\tfrac{nd}{\ell}\big)$.
\smallskip
\item[b)]
  Given $p\in\Field[X]$ of degree $d$ with $p(0)\not=0$ and $n\in\IN$,
  the $\ell$ most significant coefficients of $q:=1/p\bmod X^n$
  can be computed in time 
  $\calO\big(M(\ell\cdot\log\tfrac{n}{\ell})\cdot\log\tfrac{n}{\ell}\big)$
  where $d\leq\ell\leq n$.
\item[c)]
  Let $\Field$ have characteristic zero.
  Given $m\in\IN$, $p\in\Field[X]$ of degree $d$, and $n_1,\ldots,n_\ell\in\IN$,
  the coefficients of $p^m$ to the terms $X^{n_i}$, $i=1,\ldots,\ell$, 
  can be computed simultaneously in time
  $$ \calO\Big(d^2\cdot\big(\tfrac{\ell}{\sqrt{n}}+1\big)\cdot M\big(\sqrt{n}\big)\Big) $$
  where $d\leq\ell\leq n$ and $n_1,\ldots,n_{\ell}\leq n$ and $n\geq d^2$.
\item[d)]
  Given $p\in\Field[X]$ of degree $d$ with $p(0)\not=0$ and $n\in\IN$,
  the $n$-th to $(n-1+\ell)$-th coefficients of $q:=1/p$ 
  can be computed simultaneously
  in time $\calO\big(M(d)\cdot\log n+\ell d\big)$
  where $d\leq n$.
\end{enumerate}
\end{theorem}
\begin{proof}
\begin{enumerate}
\item[a)] is easy based on 
the observation that the $\ell$ top-most coefficients of
$p\cdot q$ depend only on the $\ell$ top-most coeffients of
both $p$ and $q$. More formally, using the convenient notation
of \mycite{Section~2}{Tellegen}, it holds
\begin{gather} \nonumber
\big\lfloor p\big\rfloor_{\deg(p)-\ell}\;=\;
\big\lceil\rev\big(\deg(p),p\big)\big\rceil^{\ell+1},
\qquad
\big\lceil p\cdot q\big\rceil^{\ell} \;=\; \big\lceil 
\lceil p\rceil^{\ell} \cdot \lceil q\rceil^{\ell}\big\rceil^{\ell},
\\  \text{and } \quad \label{e:Truncate} 
\rev\big(\deg(p)+\deg(q),p\cdot q\big)
\;=\;
\rev\big(\deg(p),p\big)\cdot\rev\big(\deg(q),q\big)
\end{gather}
where ~ $\rev\big(N,\sum\nolimits_{n=0}^N a_n X^n\big)
  :=\sum_{n=0}^N a_{N-n} X^n$ ~ and ~
$\big\lceil\sum\nolimits_{n=0}^{\infty} a_n X^n\big\rceil^{\ell}
  :=\sum\nolimits_{n=0}^{\ell-1} a_n X^n$, 
\qquad
$\big\lfloor\sum\nolimits_{n=0}^{\infty} a_n X^n\big\rfloor^{\ell}
  :=\sum\nolimits_{n=0}^{\infty} a_{n+\ell} X^n$.

Now calculate first $q:=p^{\ell/d}$ (w.l.o.g. $\ell/d$ integral)
of degree $\ell$ by repeated squaring within time $\calO(\ell)$;
and then obtain from that the $\ell$ top-most coefficients of $q^{nd/\ell}$
as the $\ell$ least ones of $\rev\big(q^{nd/\ell}\big)=\rev(q)^{nd/\ell}$ 
based on Equation~(\ref{e:Truncate}) and \mycite{Corollary~2.33}{ACT}
within $\calO\big(M(\ell)+\log\tfrac{nd}{\ell}\big)$.
\smallskip
\item[b)]
Consider the classical Newton iteration
\begin{equation} \label{e:Newton}
\tilde q \quad\mapsto\quad 2\tilde q\,-\, \tilde q^2\cdot p  \;\bmod\; X^{2\deg(\tilde q)}
\end{equation}
which yields a sequence of `approximations' $\tilde q_j$ of doubling
degrees such that $\tilde q\equiv q\bmod X^{\deg(\tilde q)}$.
In particular $q$ itself of degree $n$ (w.l.o.g. a power of 2)
is obtained after $J:=\log n$ iterations
with the running time governed by the cost of the polynomial 
multiplications in Equation~(\ref{e:Newton}) and thus dominated, 
due to the exponentially growing degree of $\tilde q$, 
by the last step \mycite{Section~9.1}{ACT}.

Let us 
analyze Newton's iteration backwards regarding which coefficients 
of $q=\tilde q_J$'s predecessors $\tilde q_{J-i}$ the $\ell$ 
top-most coefficients of $q$ depend on. 
To this end observe that Equation~(\ref{e:Newton}) turns 
some $\tilde q_{i}$ of degree $m$ first into the polynomial
$2\tilde q_{i}-\tilde q_{i}^2\cdot p$ of degree $2m+d$
and then cuts off its $d$ top-most coefficients in order
to obtain $\tilde q_{i+1}$ of degree $2m$. 
By Equation~(\ref{e:Truncate}),
the $k$ top-most coefficients of
$\tilde q_{J-i}$ having degree $m$
thus depend on and can be computed in
time $\calO\big(M(k)\big)$ from the $k+d$ top-most
terms of $\tilde q_{J-i-1}$ having degree $m/2$. 
In particular for 
the sought $\ell$ top-most coefficients of $q$ having
degree $n$ to be calculated efficiently,
it suffices to know the $\ell+dI$ top-most
ones of $\tilde q_{J-I}$ having degree $n/2^I$
as long as $\ell+dI\leq n/2^I$.
Since $d\leq\ell$, the algorithm may choose
$I:\lceil\log\tfrac{n}{\ell}-\loglog\tfrac{n}{\ell}\rceil$
and first perform the classical Newton iteration
from $\tilde q_0$ to $\tilde q_{J-I}$;
from this polynomial of degree 
$\calO\big(\ell\cdot\log\tfrac{n}{\ell}\big)$
continue via steps $J-(I-1)$ to $J$,
keeping at stage no.$(J-i)$ only the $\ell+di$
top-most coefficients.
This yields the claimed overall running time.
\item[d)]
Recall \aname{Leibniz' Rule} for higher derivatives of a product
\begin{equation} \label{e:Leibniz}
(f\cdot g)^{(n)} \quad=\quad
  \sum_{k=0}^n \binom{n}{k}\cdot f^{(k)}\cdot g^{(n-k)}
\enspace .
\end{equation}
Applied to $f:=p$ and $g:=1/(n!\cdot p)$, we obtain for $n\geq d\geq1$:
$$ 0 \;=\; \Big(p\cdot\frac{1}{n!\cdot p}\Big)^{(n)}
 \quad=\quad \sum_{k=0}^d \tfrac{1}{k!} \cdot p^{(k)}\cdot
  \Big(\frac{1}{(n-k)!\cdot p}\Big)^{(n-k)} $$
since $p^{(k)}\equiv0$ for $k>d=\deg(p)$. 
Evaluated at $x=0$ and normalized, this constitutes a linear recurrence 
like (\ref{e:ScalDef}) of depth $d-1$ 
with constant coefficients $a_k:=\frac{p^{(k)}(0)}{k!\cdot p(0)}$;
a recurrence \cite[p.41~Eq.(*)]{Tellegen}
for $P_n:=\big(\frac{1}{n!\cdot p}\big)^{(n)}(0)$, that is,
the $n$-th coefficient 
of $q=1/p=:\sum_{n=0}^\infty P_n X^n$.
Now use Theorem~\ref{t:Fiduccia}a).
\item[c)] 
W.l.o.g. $p(0)\not=0$, otherwise consider $p/X$.
Apply Equation~(\ref{e:Leibniz}) to $D^n p^{m+1}:=\big(p^{m+1}\big)^{(n)}$
in two ways\footnote{inspired by \cite[p.134]{Tricomi}} 
where $D:p\mapsto p'$ denotes the differential operator:
\begin{eqnarray*}
D^n p^{m+1}
&=&
D^{n-1}\big((m+1)\cdot p'\cdot p^m\big)
\;=\;
(m+1)\cdot\sum_{k=1}^d
\binom{n-1}{k-1} \cdot p^{(k)} \cdot D^{n-k} p^m \\
D^n p^{m+1}
&=&
\;D^n\big(p\cdot p^m\big)
\quad=\quad
\sum_{k=0}^d
\binom{n}{k} \cdot p^{(k)} \cdot D^{n-k} p^m
\end{eqnarray*}
because derivatives of $p$ higher than $d$ vanish.
Equating right sides yields%
\begin{equation} \label{e:Power}
p\cdot D^n p^m 
\quad=\quad
\sum_{k=1}^d
\underbrace{\Big((m+1)\binom{n-1}{k-1}-\binom{n}{k}\Big) 
\cdot p^{(k)}}_{=:a_k(n)} \cdot D^{n-k} g^m
\end{equation}
which, evaluated at $x=0$ and normalized by $p(0)$,
establishes a linear recurrence 
for $P_n:=D^n p^m(0)$, that is,
the $n$-th coefficient of $p^m$ (up to a factor
$n!$). This recurrence has depth $d$
and involves coefficients polynomial
in $n$ of $\deg(a_{k})\leq k$.
Now apply Theorem~\ref{t:Alin}d).
\qed
\end{enumerate}
\end{proof}
\begin{problem}
Does $1/p^m$, that is the concatenation of powering and inversion,
also admit fast partial computation?
\end{problem}
The related question concerning the product of powered and inverted
polynomials is the subject of the following section:
\section{Closure Properties}
Classical algorithms for fast polynomial arithmetic have
all significant coefficients as input and output; they are
thus obviously closed under composition and can be combined
to solve more advanced problems \cite{Shoup}. For
\emph{partial} polynomial arithmetic, on the other
hand, the output of two algorithms calculating
few coefficients of two respective high-degree polynomials
$p$ and $q$
cannot simply be fed into a third algorithm in order to
obtain merely one coefficient of, say, the product $p\cdot q$.
Instead, we refer to the framework of 
\subsection{Holonomic Functions and Recurrences}
\begin{definition} \label{d:Holonomic}
A function $f(x)$ of one variable $x$ is \emph{holonomic} 
of depth $k$ 
if it satisfies a linear ordinary differential equation 
of order $k$ 
$$  a_0(x)\cdot f^{(k)}(x)\,+\,
   a_1(x)\cdot f^{(k-1)}(x)\,+\,
  \cdots\,+\,a_{k-1}(x)\cdot f'(x)\,+\,a_k(x)\cdot f(x)\;=\; 0 
\quad\forall x $$
where the $a_i$ are requred to be polynomials.

A sequence $(P_n)_{_n}$ in the field $\Field$
is \emph{holonomic}
of depth $k$ and degree $d$
if it satisfies a linear recurrence
\begin{equation} \label{e:Holonomic}
a_0(n)\cdot P_{n+k} \,+\, a_{1}(n)\cdot P_{n+k-1} \,+\, \cdots \,+\,
a_{k-1}(n)\cdot P_{n+1}\,+\, a_k(n)\cdot P_n\;=\; 0  
\end{equation}
for all $n\in\IN$
where $a_i\in\Field[N]$ must be polynomials of degree at most $d$.
\end{definition}
By Theorem~\ref{t:Alin}, holonomic sequences admit multi-evaluation
in roughly radical time. This was exploited in 
Theorem~\ref{t:Power}c+d) whose proof reveals the following 
\begin{example}  \label{x:Holonomic}
Let $p\in\Field[X]$ denote a polynomial of degree $d$ with $p(0)\not=0$.
\begin{enumerate}
\item[a)]
 The sequence of coefficients of $1/p$ is
 holonomic of depth $d$ and degree $0$ (i.e., with constant coefficients).
\item[b)]
 For arbitrary $n\in\IN$, the (finite) sequence of
 coefficients of $p^n$ is holonomic
 of depth $d+1$ and degree $d$.
\end{enumerate}
\end{example}
It is known that a power series represents a holonomic function 
iff its coefficients form a holonomic sequence;
see e.g. \cite[p.3]{Holonomic}.
The vast and important classes of hypergeometric 
\mycite{Section~5.5}{Graham}
and generalized hypergeometric functions \cite{Nikiforov}
for instance strictly include the holonomic ones. 
Theorem~\ref{t:Alin} also yields a roughly quadratic
acceleration for their approximation:
\begin{corollary} \label{c:Holonomic}
Fix a real or complex holonomic power series $f(x)=\sum_{n=0}^\infty c_n x^n$.
Then the polynomial given by $f$'s first $N$ terms, that is,
$p_N(x):=\sum_{n=0}^{N-1} c_n x^n$,
can be evaluated at given $x$ using $\calO\big(\sqrt{N}\cdot\log N\big)$
arithmetic operations.

In particular suppose $f$ has radius of convergence $R=1/\limsup_{n\to\infty}\sqrt[n]{|c_n|}>0$ 
and fix $0<r<R$; then, given
$|x|\leq r$, one can approximate $f(x)$ up to prescribed absolute error $\epsilon>0$ 
within $\calO\big(\sqrt{\log\tfrac{1}{\epsilon}}\cdot\loglog\tfrac{1}{\epsilon}\big)$
steps.
\end{corollary}
\begin{proof}
Let (\ref{e:Holonomic}) denote the holonomic recurrence
satisfied by $(c_n)_{_n}$; for simplicity with
leading coefficient $a_0\equiv1$ --- otherwise
rescale as in the proof of Lemma~\ref{l:Polynomial}.
Then the sequence of values $\big(p_N(x)\big)_N$ satisfies
$$ \left(\begin{array}{c} p_{N}(x) \\ c_{N+1} \\ c_{N} \\
\vdots \\ c_{N-k+1} \end{array}\right) 
 \quad=\quad \left(\begin{array}{ccccc}
1 & x & 0 & \ldots & 0 \\
0 & a_{k-1} & 0 & \ldots & a_0 \\
0 &  1  & 0 & \ldots & 0 \\
0 &  0  & 1 & \ldots & 0 \\[-0.7ex]
\vdots &  \vdots  &  & \ddots & \vdots  \\
0 &  0  & 0 &  \ldots & 1 
\end{array} \right) \;\cdot\;
\left(\begin{array}{c} p_{N-1}(x) \\ c_N \\ c_{N-1} \\
\vdots \\ c_{N-k} \end{array}\right) $$
that is, a recurrence of the form (\ref{e:VecDef})
and thus supporting evaluation in the claimed time
by virtue of Theorem~\ref{t:Alin}. To choose $N$,
fix $\rho\in(r,R)$. Then $|c_n|\leq M\cdot\rho^{-n}$ for all $n$
with some appropriate $M\in\IN$ --- \textsf{Cauchy's Estimate}.
Thus $f(x)$ differs from $p_N$ by at most
$$ \sum_{n\geq N} M\cdot\bigg(\frac{|x|}{\rho}\bigg)^n
\quad=\quad M\cdot\bigg(\frac{|x|}{\rho}\bigg)^N\cdot\frac{1}{1-\tfrac{|x|}{\rho}}$$
which drops below $\epsilon$ for some $N\leq\calO\big(\log\tfrac{1}{\epsilon}\big)$.
\qed\end{proof}
\subsection{Product of Fast Partially Computable Polynomials}
The class of holonomic sequences is closed under
addition, multiplication, and convolution
\mycite{Theorem~2.1}{GFUN}. Careful inspection of
the latter proofs reveals bounds not only on the 
resulting depth but also on its degree. 
\begin{proposition} \label{p:Closure}
Let $(P_n)$ and $(Q_n)$ denote two holonomic sequences
of degree $d$ and depths $k$ and $\ell$, respectively.
Then
\begin{enumerate}
\item[a)] their sum $(P_n\!+\!Q_n)_{_n}$ is holonomic of 
 depth $K\!\leq\! k+\ell$ and degree $D\!\leq\!(k+\ell)^2d$;%
\item[b)] their product ~$(P_n\cdot Q_n)_{_n}$~ is holonomic
 of depth $K\leq k\ell$ and degree $D\leq k^2\ell^2d$;
\item[c)] their convolution 
  ~$\big(\sum_{m\leq n} P_m\cdot Q_{n-m}\big)_{n}$~
  is holonomic
 of depth $K\leq k\cdot\ell$ and degree $D\leq k^2\ell^2d$.
\end{enumerate}
\end{proposition}
As the degree of the resulting holonomic equations 
is closely related to the running time of the Gaussian
Elimination as the most expensive component in the
algorithms in \mycite{Section~2.1}{GFUN},
upper bounds on the complexity of the latter emerge
by consequence of Proposition~\ref{p:Closure}.
Our present interest however is closure under multiplication 
of fast partially computable polynomials:
\begin{corollary} \label{c:Closure}
Let $\Field$ have characteristic 0.
\begin{enumerate}
\item[a)]
 Given $p,q\in\Field[X]$ of degrees at most $d$
 with $q(0)\not=0$
 and given $m\in\IN$, one can compute
 $\ell$ arbitrary coefficients of ~$p^{m}\cdot 1/q$~
 within $$\calO\Big(d^2\cdot M(\sqrt{n d^3})+d^2\cdot\ell\cdot
  \tfrac{M(n d^3/\ell)}{n d^3/\ell}\Big)$$ operations over $\Field$.
\item[b)]
 Given $p_1,p_2\in\Field[X]$ of degrees at most $d$
 and given $m_1,m_2\in\IN$, one can compute
 $\ell$ arbitrary coefficients of ~$p_1^{m_1}\cdot p_2^{m_2}$~
 within $$\calO\Big(d^2\cdot M(\sqrt{n d^5})+d^2\cdot\ell\cdot
  \tfrac{M(n d^5/\ell)}{n d^5/\ell}\Big)$$ operations over $\Field$.
\end{enumerate}
\end{corollary}
\begin{problem}
Theorem~\ref{t:Alin}d) yields improved multi-evaluation
of holonomic recurrences with coefficients
of \emph{decaying} degrees $\deg(a_j)\leq j$.
We have already applied that in Theorem~\ref{t:Power}c)
for the partial computation of $p^m$ based on
Equation~(\ref{e:Power}). This raises the question
whether also the product/convolution of two
holonomic recurrences with decaying degrees
is again one of decaying degree. 
\end{problem}
\begin{proof}[Corollary~\ref{c:Closure}]
Since multiplication of polynomials corresponds to 
the convolution of their coefficient sequences,
combine Proposition~\ref{p:Closure}c) with
Example~\ref{x:Holonomic} to see that
the coefficients of $p^{m}\cdot 1/q$ and $p_1^{m_1}\cdot p_2^{m_2}$
form holonomic sequences of depth $D\leq\calO(d^2)$ and
degrees $K\leq\calO(d^3)$ and $K\leq\calO(d^5)$, respectively.
Then apply Theorem~\ref{t:Alin}c).
\qed
\end{proof}
The proof of Proposition~\ref{p:Closure} uses the
following extension of Observation~\ref{o:DegPattern}:
\begin{observation} \label{o:Gaussian}
\begin{enumerate}
\item[a)]
Let $A=(a_{ij})\in\Field[X]^{k\times k}$ denote a regular 
$k\times k$--matrix 
of rational functions in one variable $X$ with both
numerator and denominator of $a_{ij}(X)$ 
polynomials of degree at most $d$. 
Then the rational functions which $A^{-1}$ consists
of have degree at most $dk^2$.
\item[b)]
If furthermore the denominators in $A$ are identical 
for each column, that is, 
$a_{ij}(X)=p_{ij}(X)/q_{j}(X)$;
then $A^{-1}$ has degree at most $dk$.
\end{enumerate}
\noindent
The entries of $A^{-1}$ can in fact be
achieved to have a common denominator
while still observing the above degree bounds.
\end{observation}
\begin{proof}[Observation~\ref{o:Gaussian}]
By multiplying each column with the denominators it contains,
Claim~a) immediately reduces with $\tilde d=dk$ to b). 
For the latter, 
exploit $k$-linearity of the determinant in order to
obtain $\prod_{j=1}^k q_j(X)$ of degree$\leq dk$
as the denominator of $\det(A)$ with numerator 
~ $ P \;:=\; \sum_{\sigma\in{\mathcal S}_k} 
  \sgn(\sigma)\prod_{j=1}^k p_{\sigma(j),j} $
of degree at most $dk$ as well.
Now $A^{-1}$ has entries $\det(A_{ji})/\det(A)$
based on Cramer's Rule with $A_{ij}$ a sub-matrix
of $A$ and its determinant thus a rational function
of degree at most $d(k-1)$.
More precisely, the denominator of $\det(A_{i,j_0})$
is $\prod_{j\not=j_0} q_j(X)$ and thus cancels out
in $\det(A_{ij})/\det(A)$ against the denominator 
of $\det(A)$, leaving $P$ as common denominator 
of all entries in $A^{-1}$.
\qed\end{proof}
\begin{proof}[Proposition~\ref{p:Closure}]
By prerequisite, the $k$--shifted sequence
$(P_{k+n})_{_n}$ is 
a linear combination of the original (i.e.,
0--shifted), the 1--shifted, \ldots, and
$(k-1)$--shifted one; a linear combination
with coefficients being rational functions 
of degree at most $d$
and common denominator. By induction on $m$, also
the $(k+m)$--shifted sequence is
a linear combination of the first $k$ shifts --- this
time with coefficients of degree at most $md$ and common
denominator. 

In particular, the vector space $U$ (over
the field $\Field(N)$ of rational functions in $N$) 
formed by all shifts of $(P_n)_{_n}$
has dimension at most $k$; similarly, the shifts
of $(Q_n)_{_n}$ give rise to a vector space
$V$ of dimension at most $\ell$.
Therefore, the vector space
$U+V$ of all joint shifts is at most
$(k+\ell)$--dimensional, that is, 
latest the $(k+\ell)$--shift of $(P_n+Q_n)_{_n}$
is a linear combination of its
predecessors: closure of holonomic
sequences under addition at depth at most
$k+\ell$.

In order to estimate the degree of the
rational coefficients involved in the latter
linear combination, express each of the first 
$k+\ell+1$ shifts of $(P_n+Q_n)_{_n}$ 
as linear combinations of the first $k$ shifts
of $(P_n)_{_n}$ and the first $\ell$ shifts
of $(Q_n)_{_n}$. By the above remark, this
gives rise to a $(k+\ell)\times(k+\ell+1)$--matrix
$B$ over $\Field(N)$ with entries of degree at
most $d\cdot\max\{k,\ell\}$ and in each column at most two
different denominators --- which is easy to turn
into degree $D\leq d\cdot(k+\ell)$ with column-wise single
common denominators. The $k+\ell+1$ columns of $B$ are linearly
dependent, either by the above considerations
or simply due to its format. 

Suppose for simplicity that 
the first $K:=k+\ell$ columns are independent ---
otherwise we argue similarly to obtain an even
shorter and lower-degree recurrence.
Expressing the $(k+\ell+1)$--st column by 
these first ones yields an explicit representation 
of the $(k+\ell)$--shift of $(P_n+Q_n)_{_n}$ in terms
of its first $k+\ell$ shifts. Denoting by
$b$ the last column of $B$ and by $A$ its
first $k+\ell$ columns, we obtain 
as the coefficients of this representation
the vector $A^{-1}\cdot b$ which, by virtue
of Observation~\ref{o:Gaussian}b), consists
of rational functions of degree 
$\calO(DK)=\calO\big(d(k+\ell)^2\big)$ with
common denominator. We have thus arrived at the 
desired holonomic recurrence 
(\ref{e:Holonomic}) for $(P_n+Q_n)_{_n}$.

For $(P_n\cdot Q_n)_{_n}$, consider the tensor product 
vector space $U\otimes V$ of dimension at
most $k\cdot\ell$ to obtain a recurrence of the
claimed depth. A generator of $U\otimes V$ is the
collection of mixedly-shifted product sequences
$(P_{n+i}\cdot Q_{n+j})_{_n}$ with $0\leq i<k$
and $0\leq j<\ell$. Therefore, each of the
$K:=k\cdot\ell$ (singly but farther) shifted sequences
$(P_{n+m}\cdot Q_{n+m})_{_n}$, $0\leq m\leq K$,
can be expressed as a linear combination of
this generator; in fact with coefficients
being rational functions of degree $D\leq dK$
with common denominator.
Putting them into a $(K+1)\times K$--matrix
and arguing as above, we obtain a representation
of the $K$-shifted sequence $(P_{n+K}\cdot Q_{n+K})_{_n}$
as linear combination of the $m$-shifts, $0\leq m<K$,
with coefficients being rational functions of
degree $\calO(DK)=\calO(dk^2\ell^2)$.

The proof for convolution proceeds similarly.
\qed\end{proof}

\end{document}